\documentclass[12pt]{iopart}

\usepackage{amssymb,amscd,epsfig,bbold,stmaryrd,bm,dsfont}
\usepackage{graphicx}
\usepackage{color}
\usepackage{psfrag}
\usepackage{iopams}
\usepackage{dsfont}
\usepackage{setstack}
\usepackage{bm}

\begin{document}

\title[Dephasing of fluxonium qubit]{Dephasing due to quasiparticle tunneling in fluxonium qubits: a
phenomenological approach}

\author{Samuele Spilla}
\address{Dipartimento di Fisica e Chimica, Universit\`a di Palermo, I-90123 Palermo, Italy}
\address{Institut f\"ur Theorie der Statistischen Physik, RWTH Aachen University, D-52056 Aachen, Germany}

\author{Fabian Hassler}
\address{JARA-Institute for Quantum Information, RWTH Aachen University, D-52074 Aachen, Germany}

\author{Anna Napoli}
\address{Dipartimento di Fisica e Chimica, Universit\`a di Palermo, I-90123 Palermo, Italy}

\author{Janine Splettstoesser}
\address{Department of Microtechnology and Nanoscience (MC2), Chalmers University of Technology, SE-41298 G\"{o}teborg, Sweden}

\begin{abstract} 
The fluxonium qubit has arisen as one of the most promising candidate devices
for implementing quantum information in superconducting devices, since it is
both insensitive to charge noise (like flux qubits) and insensitive to flux
noise (like charge qubits).  Here, we investigate the stability of the quantum
information to quasiparticle tunneling through a Josephson junction.
Microscopically, this dephasing is due to the dependence of the quasiparticle
transmission probability  on the qubit state. We
  argue that on a phenomenological level the dephasing mechanism can be
understood as originating from heat currents, which are flowing in the device
due to possible effective temperature gradients, and their sensitivity to the
qubit state. The emerging dephasing time is found to be insensitive to the
number of junctions with which the superinductance of the fluxonium qubit is
realised.  Furthermore, we find that the dephasing time increases
quadratically with the shunt-inductance of the circuit which highlights the
stability of the device to this dephasing mechanism.
\end{abstract}

\maketitle

\section{Introduction}

Among the various types of superconducting
qubits~\cite{Makhlin01,You05,Clarke08}, the recently developed fluxonium
qubit~\cite{Koch09,Manucharyan09} has the unique advantage of being protected
against  both charge and flux noise.  This is important since both effects in
general limit the performance of the qubits  by
introducing relaxation and dephasing processes. Indeed, over the last few
years considerable effort has been made in order to understand, and
consequently to reduce, the causes of relaxation and decoherence in different
types of superconducting
circuits~\cite{Makhlin01,Martinis05,Paladino10,Skinner08,HerreraMarti13}.

Initially, the fluxonium qubit has been designed in order to reduce the
sensitivity of the Cooper pair box to charge noise
\cite{Koch09,Manucharyan09}.  Subsequently, it has been argued that it is also
insensitive to flux noise \cite{HerreraMarti13}.
 In order to reach a regime in which these relaxation and decoherence
 processes are exponentially suppressed, the charging energy $E_C = e^2/2C$,
 with the capacitance  $C$ and the elementary charge $e$, has to be
much larger than the inductive scale $E_L = e^2/4 \alpha^2 L$, with  the
inductance $L$ and  the fine-structure constant $\alpha=e^2/\hbar c$. While it
is impossible to realize such large ``superinductances", $L\gg \alpha^{-2} C
\simeq 10^4 C$, with conventional media, in the fluxonium qubit it has been
realized by an array of (large) Josephson junctions~\cite{Manucharyan09}.  A
downside of this approach is that the device is then potentially plagued by
spurious phase slips through the array; however, these have been successfully
eliminated \cite{Manucharyan12,Masluk12}.  Recently, different detrimental effects in
the fluxonium qubit due to non-equilibrium quasiparticles have been addressed
theoretically \cite{Catelani11a,Catelani11,Catelani12,Leppakangas12,Zanker15}
and experimentally \cite{Pop14}.

Of all the processes limiting the performance of the qubits discussed above
quasiparticle tunneling is particularly important as it is \emph{intrinsic} to the
superconducting tunneling junction and as such forms an absolute limit.  While
relaxation and dephasing mechanisms due to quasiparticle tunneling have been
extensively studied in charge qubits
\cite{Leppakangas11,Lutchyn05,Lutchyn06,Catelani14,riste:12,Wang14}, for the flux
qubits most studies so far have concentrated on the relaxation due to
quasiparticle processes \cite{Catelani11a,Catelani11,Pop14}. On the other hand, first investigations of the dephasing due
to quasiparticle tunneling on a perturbative level in the junction transmission have been put forward in
\cite{Catelani12,Leppakangas12,Zanker15}. It turns out that treating the
problem in perturbation theory leads to a diverging  result due to the sharp
peak of the quasiparticle density of states at the gap. This problem has been
addressed by introducing as a cutoff a relaxation rate \cite{Leppakangas12} or a
dephasing rate \cite{Catelani12,Zanker15} that broaden the density of states, 
where the latter has been determined self-consistently. It has been discussed
that in principle the divergence is lifted by treating the tunneling
nonperturbatively as the divergence simply signals the presence of a weakly
bound Andreev state close to the gap \cite{Catelani12}.

Recently, it has been recognised that the presence of different nonequilibrium
quasiparticle distributions on the different superconducting islands  of the
qubit (possibly accidentally arising during operation of the qubit~\cite{riste:12} and resulting in an effective \emph{temperature gradient} of stationary nonequilibrium quasiparticle distributions) can lead to an additional
decoherence mechanism for the flux qubits, see reference \cite{Spilla14},
where in particular the flux qubit in the Delft-qubit design has been
addressed.  It has been shown that in this case a limitation of the dephasing
time arises caused by heat currents carried by 
quasiparticles which flow through the Josephson junctions of a superconducting
qubit as a response to the (effective) temperature gradient.  

The microscopic origin of this dephasing mechanism is the fact that a heat
current flowing through a Josephson junction depends on the phase difference
of the two superconductors separated by the junction.  This fundamental effect
has been predicted over 50 years ago, see \cite{Maki65}, and has later been
studied in more detail in references \cite{Guttman97,Guttman98,Zhao03,Zhao04}.
Only very recently, the phase dependence of heat currents through Josephson
junctions has been measured experimentally
\cite{Giazotto12a,Giazotto12b,Martinez14}.  Intriguingly, due to the general
phase-dependence of the heat current, heat currents flowing across junctions
of superconducting flux qubits can depend on the qubit state.  This results in
the dephasing of the qubits, since heat currents are dissipative
\cite{Spilla14}.

In this paper, we investigate pure dephasing of the fluxonium qubit due  to a
non-equilibrium quasiparticle distribution.  We in particular analyze the
relevance of this intrinsic dephasing mechanism (without population
relaxation) slightly away from the sweet spot, where it would be suppressed, see \cite{Catelani11,Catelani12,Leppakangas12}.  We model the
non-equilibrium distribution of quasiparticles by an effective temperature
which is different on different superconducting segments of the qubit. We study in details the effect of the
heat current flowing both through the Josephson junction of the so-called
\emph{black-sheep} junction with Josephson energy $E_\mathrm{J}$, which is the
main source of nonlinearity, as well as through the array of larger junctions
constituting the superinductance, which is shunting the \emph{black sheep}. In
particular, we are interested in the linear response regime where the
difference in temperatures is small. We show that the sensitivity of the fluxonium qubit to heat transport and the resulting pure dephasing is suppressed by a factor $E_L^2/E_\mathrm{J}^2$ (up to logarithmic corrections) rendering the fluxonium qubit rather insensitive to
this dephasing source.
We point out that even if the effective temperature gradient is vanishingly small, the study of the heat conductance (rather than of the heat currents) is relevant: intriguingly, it can be used as a phenomenological approach for the investigation of dephasing due to nonequilibrium quasiparticles.\footnote{Note that a study of the charge current (or conductance) would not be appropriate
here, since the charge of quasiparticles depends on their composition of
electron- and hole-like states. In particular, this has as a consequence that the charge current is governed by the group velocity of electrons and holes, which vanishes at the gap, see e.g.~\cite{Blonder82}. This hinders the access to the phase-dependent transmission close to the gap, which is important for quasiparticle dephasing, via the charge current.} Different from our phenomenological approach,
prior work on dephasing in flux-based qubits
\cite{Catelani11,Catelani12,Leppakangas12,Zanker15} was based on a microscopic
model.  Our approach has the advantage that we are able to consider very small
temperature gradients -- corresponding to superconducting qubit segments with
identical superconducting gaps. In this regime, the dominating effect of the heat-current sensitivity to the qubit state stems from the phase-dependence of a weak bound state originating from Andreev reflection. Due to the fact that our
approach is \emph{nonperturbative} in the tunnelling coupling, we are not
plagued by divergencies as the weak Andreev bound state acts as a natural
phase-dependent cut-off. Additionally, our approach provides a nice link of
this intrinsic dephasing mechanism to thermal transport quantities and their
ability to ``measure" a qubit state.

The paper is organized as follows: In section~\ref{fluxonium}, we introduce
the generic Hamiltonian of the fluxonium qubit.  We recall the results for
heat currents flowing through a single Josephson junctions for very small
temperature gradients in section~\ref{heatcurrent}.  These findings are then
used in section~\ref{fluxonium_heat} to investigate on the sensitivity of the
heat current on the fluxonium states.  In section~\ref{sec_dephasing}, we show
in how far the resulting dephasing can limit the operation of the fluxonium
qubit.

\section{Fluxonium qubit}\label{fluxonium}

\begin{figure}[tb]
\centering
\includegraphics[width=.95\linewidth]{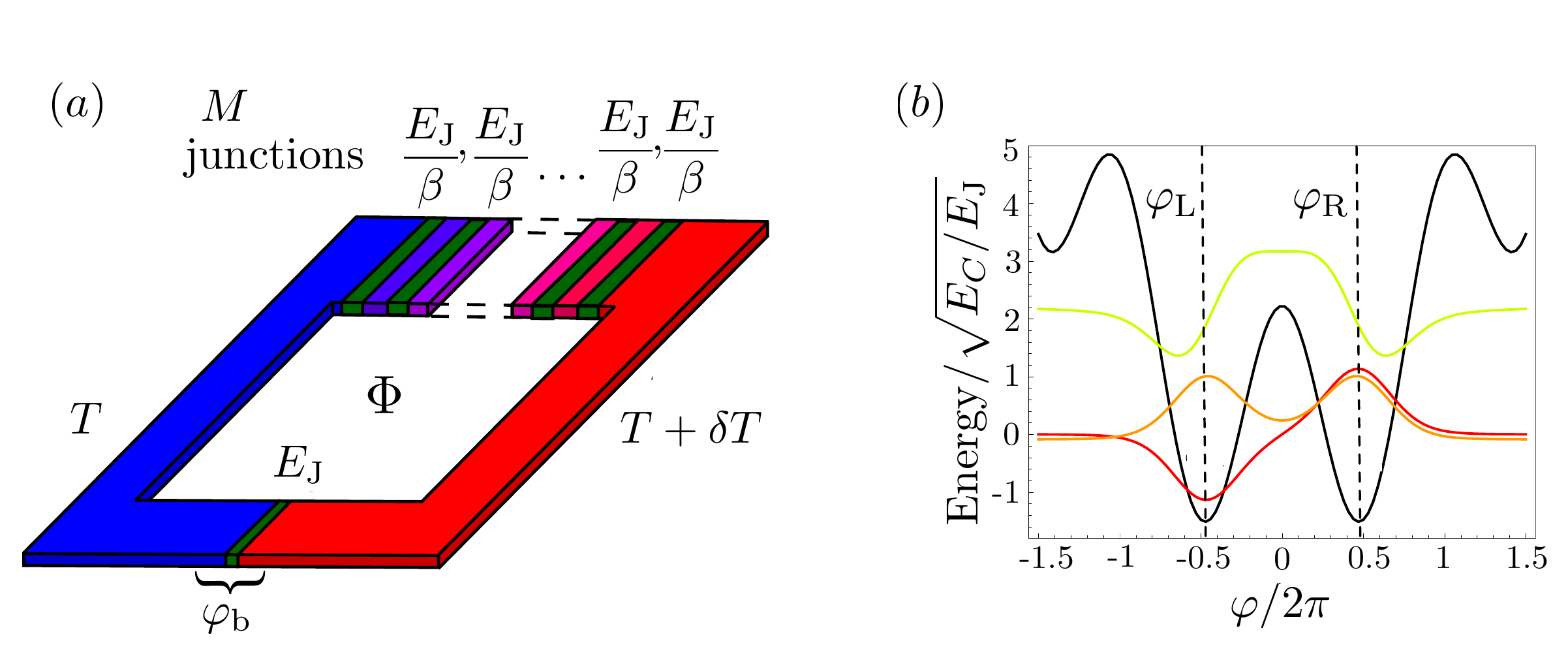}
\caption{(a) Sketch of a fluxonium qubit, consisting of a
superconducting loop interrupted by $M+1$ Josephson junctions, which is threaded by a magnetic flux $\Phi \approx
  \Phi_0/2$.  The \emph{black sheep} junction has a Josephson energy $E_\mathrm{J}$ and a phase difference
  $\varphi_\mathrm{b}$; the remaining $M$ junctions with a larger Josephson energy $E_\mathrm{J}/\beta$ ($\beta <1$)
  act as a superinductance.  We consider the situation where the two
  electrodes separated by the \emph{black sheep} are biased by a small
  temperature difference. (b) Potential energy of the fluxonium as a function of the phase $\varphi$ for
  $E_{L}/E_\mathrm{J}= 6 \times 10^{-2}$ at the sweet spot, $f=1/2$ \cite{Manucharyan09}. The wave functions of the three lowest lying eigenstates
  ($E_{C}/E_\mathrm{J}=3 \times 10^{-1}$) are depicted by red, orange and green
  lines, where the vertical offset indicates the corresponding eigenenergy.
} \label{fig:fluxonium}
\end{figure}

The fluxonium is a superconducting qubit which can be thought of as a
Cooper-pair box inductively shunted with a superinductance \cite{Koch09}.
Its electrical circuit is shown in figure~\ref{fig:fluxonium}(a): it consists of a
Josephson junction, which we refer to as the \textit{black
sheep}, with Josephson energy $E_\mathrm{J}$ and charging energy $E_C$.  This
Josephson junction is shunted by an array of $M$ larger Josephson junctions
(with Josephson coupling energy $E_\mathrm{J}/\beta$, $\beta < 1$).  When
operated at microwave frequencies well below its self-resonant frequency
$\sqrt{E_\mathrm{J} E_{C}/\beta}/\hbar$, this array emulates a
`superinductance' for sufficiently large $M$.  It is the presence of this
superinductance which renders the fluxonium insensitive to charge noise while
yielding a highly anharmonic spectrum, both important points for the
realisation of a qubit.

\subsection{Hamiltonian}
The Hamiltonian of the fluxonium consists of different parts, stemming from the charging and the Josephson energy of the different junctions~\cite{Masluk,Ferguson13}.  Charging
effects on the \textit{black sheep} junction lead to the charging energy $T =
4 E_{C}\hat{n}_\mathrm{b}^2$ with $\hat{n}_\mathrm{b}$ the number of
Cooper pairs on the capacitance plate of the \textit{black sheep}.  In the Hamiltonian, the
charging energy of the \textit{black sheep} serves as the kinetic energy, while the sum of the different Josephson energies takes the role of the potential energy.  In particular, the potential energy of the \textit{black
sheep} is given by $U_\mathrm{b}=E_\mathrm{J}(1-\cos\varphi_\mathrm{b})$
with $\varphi_\mathrm{b}$ the superconducting phase difference across the
\textit{black-sheep} junction which is the conjugate variable to
$\hat{n}_\mathrm{b} = -i \partial/\partial \varphi_\mathrm{b}$.  For
simplicity, we assume the junctions in the array constituting the
superinductance to be equal with a somewhat larger Josephson energy
$E_\mathrm{J}/\beta$, $\beta <1$.  The total
potential energy is thus given by $U = U_\mathrm{b} + M (E_\mathrm{J}/\beta)
(1- \cos \varphi_M)$ with $\varphi_M$ the phase difference across each of the
junctions constituting the superinductance. We neglect the capacitive
energies of the M array junctions since their larger area translates into
smaller charging effects. The radius $r$ of the
superconducting loop is supposed to be so small that the inductive energy
$\Phi_0^2/r$ per flux quantum $\Phi_0 =hc/2e =\pi e/\alpha$ dominates the
Josephson energies $E_\mathrm{J}$.  In this regime, the magnetic flux in the
superconducting loop is quantized which leads to the condition
$\varphi_\mathrm{b}+M\varphi_M+2\pi f=0 \; (\mathrm{mod} \,2\pi)$ with the
dimensionless parameter $f=\Phi/\Phi_0$ due to the magnetic flux $\Phi$
accounting for the external magnetic field penetrating the loop.  Using this
relation, we can express the Hamiltonian in terms of one dynamical variable, only, which we choose
to be the total phase difference across the superinductance, $\varphi = M \varphi_M = -\varphi_\mathrm{b} - 2\pi f$.  
Note that at
fixed $\varphi$, the phase difference $\varphi_M$ across each junction in the
array constituting the superinductance becomes small for large $M$.
As a result of the small phase difference across each of the large junctions~
\footnote{We here neglect potential phase slips across the array, which is justified by the large array capacitance.}, we are allowed to expand the cosine in
the potential energy to second order in $\varphi_M$ with the result 
\begin{equation}\label{eq:expansion}
U=-E_\mathrm{J}\cos{\varphi_\mathrm{b}}+ \frac12 E_{L} \varphi^2;
\end{equation}
(up to an irrelevant constant shift). Here, we have introduced the inductive energy of the superinductance
\begin {equation}\label{EL}
  E_{L}=\frac{E_\mathrm{J}}{\beta M}=\frac{(\Phi_0/2\pi)^2}{L_\mathrm{eff}}.
\end{equation}
with the effective inductance $L_\mathrm{eff}$, which increases linearly with $M$.
It is the inductive term, which we obtained from expanding the cosine potential of the array to second order
in $\varphi_M$, which breaks the $2\pi$ periodicity of the total potential
$U$ as a function of $\varphi$, thus rendering the device charge insensitive
\cite{Koch09,Ferguson13}.

In conclusion, the total Hamiltonian $H= T + U$ of the fluxonium is given by
\begin{equation}\label{Hamiltonian}
H=
4E_{C}\hat{n}^2 +\frac{1}{2}E_{L}\varphi^2- E_\mathrm{J}\cos (\varphi+2\pi f )
\end{equation}
with $\hat{n} = -i\partial/\partial \varphi = -\hat{n}_\mathrm{b}$ the
operator conjugate to $\varphi$.  The Hamiltonian (\ref{Hamiltonian}) for
$f\approx 1/2$ may serve as a model-Hamiltonian for different qubit types
\cite{Masluk}.  In the present manuscript, we mainly discuss the
fluxonium qubit which is realized in the regime $E_\mathrm{J} \gg E_C \gg E_L$, where the latter inequality is made possible by the presence of the superinductance.  We compare the result to the case of
the Delft qubit realized in the regime $E_L \simeq E_\mathrm{J} \gg E_C $; see
table~\ref{table} for a discussion of the different energy scales.

\begin{table} \begin{center}
    \begin{tabular}{ c  c c } \hline
       & $E_{L}/E_\mathrm{J}$ & $E_{C}/E_\mathrm{J}$ \\ \hline Fluxonium &
  $6\times 10^{-2}$ & $3\times 10^{-1}$ \\ Delft qubit & $ 5\times 10^{-1}$ &
  $2\times 10^{-2}$\\ \hline\hline
\end{tabular} \end{center}
 \caption{ \label{table}
 Characteristic energy ratios for the fluxonium
 \cite{Manucharyan09} and Delft qubit \cite{Mooij99} when described by the
 Hamiltonian (\ref{Hamiltonian}) for the dynamic degree of freedom.  The ratio
 $E_\mathrm{J}/E_C$ approximately corresponds to the number of levels in a potential minimum, whereas $E_\mathrm{J}/E_L$ is a measure of the number of mimima of the potential which typically contribute to the qubit states \cite{Masluk}.}
\end{table}

\subsection{Qubit states}\label{qubit_states}

Close to $f = 1/2$, the potential landscape is given by a double-well
potential, which is an ideal starting ground to encode a qubit, see figure
\ref{fig:fluxonium}(b).  For the fluxonium qubit, we find numerically that the
potential landscape features two well-localized minima for $|\delta f|\lesssim
0.3$ with $\delta f=f-1/2$.  The two minima are situated at $\varphi_\mathrm{L/R}$.  The two lowest-lying states in these minima, representing the qubit states, correspond to a current flowing
clockwise/counter-clockwise in the device.  However, due to the large
inductance the
corresponding currents are rather small, even though the two states differ by a large magnetic flux. 

The dephasing mechanism which we investigate in the following arises due to the fact that the
different semiclassical states correspond to different superconducting phases
on the islands which in turn makes the phase-dependent heat currents through the junctions dependent on the
qubit state.  The resulting decoherence thus projects the qubit on the
semiclassical states $\varphi_\mathrm{L/R}$. 

We now exactly look at these semiclassical states.
The position of the minima $\varphi_\mathrm{L/R}$ is given by the
solution  $\bar\varphi$ to the equation
\begin{equation}\label{eq:minima}
  0 = \frac{\partial U}{\partial\varphi} = E_L \bar\varphi + E_\mathrm{J}
  \sin(\bar\varphi + 2\pi f).
\end{equation}
At $\delta f= 0$, the solutions are situated
symmetrically around $\varphi=0$ with $\varphi_\mathrm{R} = -\varphi_\mathrm{L} = \varphi^*$. In general, this transcendental equation can only be solved numerically.  Therefore, in the following, we present all analytical results in the regime $E_L \ll E_\mathrm{J}$,
whereas the numerical results presented in the figures are obtained taking into account the exact solution to equation
\ref{eq:minima}.  
In the fluxonium limit $E_L \ll E_\mathrm{J}$, we have $\varphi^*\approx \pi ( 1 -
E_L/E_\mathrm{J})$.  For small $\delta f\neq0$ the minima shift slightly to the right and
are given by $\varphi_\mathrm{R} \approx \varphi^* - 2\pi (1- \pi E_L/E_\mathrm{J}) \delta f $
and $\varphi_\mathrm{L} \approx -\varphi^* - 2\pi (1-\pi E_L/E_\mathrm{J}) \delta f$.

In the vicinity of a local minimum $\bar{\varphi}=\varphi_\mathrm{L/R}$ of the
potential shown in figure~\ref{fig:fluxonium}(b), the potential energy can be
approximated by a harmonic oscillator potential, $U(\varphi)\approx U(\bar{\varphi})+
\frac{1}{2} E_\mathrm{J}(\varphi-\bar{\varphi})^2$ with the lowest-lying
states corresponding to the ground state wave-functions in each of the
potential minima.  The spread $\delta \varphi$ of the wavefunction is given by
$\delta \varphi \simeq \sqrt{E_C/E_\mathrm{J}}$.  As both the fluxonium qubit and the
Delft qubit are in the semiclassical limit, $E_C \ll E_\mathrm{J}$, we neglect the
finite extent of the wave functions in the following and assume that they are well
localized at the single value $\varphi_\mathrm{L/R}$ of the phase variable. With
that our results become independent of $E_C$.

\section{Heat current in a Josephson junction -- linear response regime} \label{heatcurrent}

In the following, we want to study the effect  of a small (effective) temperature gradient
on the quantum information encoded in the fluxonium. To this end, we need to
calculate the resulting heat current flowing through the junctions interrupting the
superconducting loop of the fluxonium qubit.  We are interested in the effect
of accidental (effective) temperature differences; this means that we need to calculate 
heat currents in response to small temperature differences $\delta T$, that is
$k_\mathrm{B}\delta T /\Delta\sim 10^{-2}$.
We hence evaluate the
heat current in the linear response regime.  In this section, we briefly
recall the results of \cite{Zhao03,Zhao04} for heat currents in a single Josephson
junction in the linear-response regime for small temperature gradients before
we continue by investigating the effects of the heat current on the
fluxonium qubit in the next sections.

We consider two superconducting reservoirs with gaps $\Delta_1=\Delta(T_1)$ and $\Delta_2=\Delta(T_2)$
and phase difference $\phi$, interrupted by a junction with transmission
probability $D$.  Taking the number of channels contributing to transport to
be $N$, the normal-state resistance of the junction is given by $R
=h/(2e^2ND)$.  We now assume that the two reservoirs are kept at temperatures
$T_1=T$ and $T_2=T+\delta T$.  Being interested in small temperature
gradients, $\delta T/T\ll1$, only, we have $\Delta_1=\Delta_2=\Delta$.~\footnote{Importantly, in contrast to the procedure employed here, perturbative approaches in the tunnel coupling such as used in references~\cite{Spilla14,Maki65}, need to introduce artificial cutoff energies to avoid emerging divergencies for $\Delta_1\approx\Delta_2$.}  In
this regime, the heat current between the two superconducting electrodes is
given by the linear response result~\cite{Zhao03,Zhao04} 
\begin{equation}\label{eq_heatgeneral}
\dot{Q}(\phi, T, \delta T)=-\kappa(\phi, T)\delta T
\end{equation}
with the thermal conductance
\begin{eqnarray}
\label{KappaLinearRegime}
\fl \kappa=\frac{1}{2e^2 Rk_\mathrm{B}T^2}\int_{\Delta}^{\infty}d\omega\,\frac{\omega^2}{
  \cosh^2{(\omega/2k_\mathrm{B}T)}}\frac{\omega^2-\Delta^2}{(\omega^2-\omega_\mathrm{c}^2)^2}\left[(\omega^2-\Delta^2\cos{\phi})-D\Delta^2\sin^2{\frac{\phi}{2}}\right].
  \end{eqnarray}
Here, we have introduced the energy $\omega_\mathrm{c}=\Delta
[1-D\sin^2(\phi/2)]^{1/2}$ of the weakly bound Andreev state emerging in the
junction \cite{beenakker}.  Most importantly, the heat current depends on the
phase difference $\phi$ across the junction.  We are interested in the
tunnelling regime with $D\ll1$, such that the last term in
(\ref{KappaLinearRegime}) can be neglected with the result
\begin{eqnarray}\label{thermal current}
\kappa=\frac{1}{2e^2R
  k_\mathrm{B}T^2}\int_{\Delta}^{\infty}d\omega\,\frac{\omega^2}{
    \cosh^2{(\omega/2k_\mathrm{B}T)}}\frac{\omega^2-\Delta^2}{(\omega^2-\omega_\mathrm{c}^2)^2}(\omega^2-\Delta^2\cos{\phi}) .
\end{eqnarray}
In the following, it is important to make the phase dependence of
equation \ref{thermal current} explicit. As outlined in \ref{app:weak}, we can
bring the expression in the form
\begin{equation}\label{zhao}
\kappa=\kappa_0-\kappa_1\sin^2{\frac{\phi}{2}}\ln{\left(\sin^2{\frac{\phi}{2}}\right)}+\kappa_2\sin^2{\frac{\phi}{2}}
\end{equation}
with the coefficients
\begin{eqnarray}\label{ks}
\fl \kappa_0=\frac{1}{2e^2 R k_\mathrm{B} T^2}\int_{\Delta}^{\infty}d\omega\,
\frac{\omega^2}{\cosh^2 (\omega/2 k_\mathrm{B} T)}, \qquad
\kappa_1=\frac{\Delta^3}{2 e^2 R k_\mathrm{B}T^2 \cosh^2 (\Delta/ 2 k_\mathrm{B} T)} 
, \nonumber\\
\fl\kappa_2=  \kappa_1 \bigl|\ln D\bigr|
 + \frac{\Delta^2}{e^2 R k_\mathrm{B} T^2} 
\int_{\sqrt{3} \Delta}^\infty
\frac{\omega^2}{\cosh^2(\omega/2  k_\mathrm{B} T) (\omega^2 -\Delta^2)}
\\
\fl\qquad+ \frac{\Delta^2}{e^2 R k_\mathrm{B} T^2} 
\int_{\Delta}^{\sqrt{3} \Delta} 
\frac{\omega}{\omega^2-\Delta^2} \left( \frac{\omega}
{\cosh^2(\omega/2 k_\mathrm{B} T)} - \frac{\Delta}
{\cosh^2(\Delta/2 k_\mathrm{B} T)}   \right)  \nonumber
\end{eqnarray}
which are independent of the phase difference. Note that, due to the specific derivation that we chose for these parameters (see \ref{app:weak}), differences in the functional form of these coefficients occur with respect to the ones given in reference~\cite{Zhao03,Zhao04}. For small temperatures compared to the critical one $T\ll T_\mathrm{c}$, we
observe that $\kappa_{0,1,2} \propto e^{-\Delta/k_\mathrm{B} T}$, i.e., they
are exponentially suppressed at low temperatures.

\section{Heat currents in the fluxonium qubit}\label{fluxonium_heat}

We now proceed to investigate how the phase sensitivity of the heat current
through a Josephson junction manifests itself in the fluxonium. The situation
we have in mind is some stationary nonequilibrium quasiparticle distribution
on the fluxonium qubit. As mentioned above, we model
this distribution by an effective temperature, in order to keep the discussion simple.~\footnote{Note that more general nonequilibrium quasiparticle distributions can be analyzed by replacing the Fermi functions (entering through the $cosh$ term in equations~\ref{KappaLinearRegime} to \ref{ks}) by arbitrary distribution functions.} In particular, we treat the
situation where one side of the \text{black-sheep} Josephson junction is at an
elevated temperature $T_2=T+\delta T$ with respect to the other which is at
temperature $T_1=T$, see figure~\ref{fig:fluxonium}(a). As a result
quasiparticles tunnel from the ``hot'' to the cold reservoir. As seen in
the last section, the resulting heat current  depends on the
superconducting phase differences and thus on the state of the qubit. As we
have assumed the $M$ Josephson junctions of the array emulating the
superinductance to be equivalent, the temperature gradient is distributed
among the $M$ elements with the temperature difference  $\delta T_M=\delta
T/M$ on a single junction.  The heat current $\dot{Q}$ flowing into the cold
reservoir, which is held at temperature $T_1=T$ (or equivalently in the heat
current flowing out of the hot reservoir, kept at temperature $T_2=T+\delta
T$~\footnote{Being interested in the stationary situation, we can make this
assumption as long as coupling to phonons is neglected.}) is given by the sum
of two terms
\begin{equation}
 \dot{Q} = \dot{Q}(-\varphi_\mathrm{b}, T,  \delta T)+\dot{Q}(\varphi_M, T, \delta
 T/M), \label{eq_totalheat}
\end{equation}
where the first term is the contribution of the \textit{black sheep} and the
second term is due to the large Josephson junction connecting the array of the
superinductance to the cold reservoir.

The decoherence of the fluxonium qubit is triggered by the difference
$\delta \dot Q = | \dot Q_\mathrm{R} - \dot Q_\mathrm{L}|$ of the heat currents flowing in the
device when the qubit is in the state R/L with $\varphi= \varphi_\mathrm{R/L}$.
Following our previous work, \cite{Spilla14},  we introduce the
sensitivity
\begin{equation}\label{s}
s=\frac{\delta \dot{Q}}{|\dot{Q}_\mathrm{R}+\dot{Q}_\mathrm{L}|} 
\end{equation}
as a measure of correlation between the heat current and the qubit state.  As a
large sensitivity corresponds to a large difference of heat currents for
different qubit states, we expect that this in turn leads to fast dephasing
of the qubit; an expectation which we confirm below.  However, in a first
step, we want to calculate the sensitivity for the fluxonium.

\subsection{Effect of the number of junctions implementing the superinductance}

In this section, we show that the sensitivity of the heat currents to the fluxonium state only depends on
the effective parameters $E_{L}$ and $E_\mathrm{J}$ of the Hamiltonian, given in equation~\ref{Hamiltonian}, and
not on the specific number  $M$ of junctions (or their asymmetry factor $\beta$)
with which the superinductance $L_\mathrm{eff}$ is emulated.
In order to realise a fluxonium qubit modelled  by the Hamiltonian given in
equation~(\ref{Hamiltonian}), a large amount $M$ of array junctions is needed,
which all have a Josephson energy larger than the one of the \textit{black
sheep} by a factor $\beta^{-1}$. In view of equation (\ref{EL}), to obtain a given value of $E_{L}$, the coupling strength of each array junction needs to scale
like $M \propto \beta^{-1}$.

It is clear that from the two terms in equation~(\ref{eq_totalheat}), only the
one stemming from the junction of the array forming the superinductance, $\dot{Q}(\varphi_M, T, \delta T/M)$,
could possibly depend on $M$. The variables occurring
in the argument of the heat current depend on $M$ via
$\varphi_M=-(\varphi_{b}+2\pi f)/M$ and $\delta T/M$. Exploiting
equations~(\ref{zhao}) and (\ref{ks}), we find that for large $M$, where
$\varphi_M\rightarrow0$, the only relevant term  for
$\dot{Q}(\varphi_M\approx0, T, \delta T/M)$ is given by the phase-independent part of the thermal conductance of the array junction, which we denote by $\kappa_{0M}(T)$.  
As a consequence, the state-dependent heat current difference $\delta\dot{Q}=|\dot{Q}_\mathrm{R}-\dot{Q}_\mathrm{L}|$ depends only on the heat current through the black sheep. Hence in the following only the dependence on $M$ of the term $\kappa_{0M}(T)$ entering the denominator of the sensitivity has to be investigated.

The thermal conductance $\kappa_{0M}$ is proportional to $R_M^{-1}$, the normal-state resistance of the outer junction of the array. Now,
due to the generalised Ambegaokar-Baratoff
relations~\cite{Ambegaokar63}, the normal-state resistance is
inversely proportional to the Josephson energy of the corresponding junction
and we thus find $\kappa_{0M} = \beta^{-1}\kappa_0$, with $\kappa_0$ the phase-independent contribution to the thermal conductance of the \emph{black sheep}.  Therefore, $\dot{Q}(\varphi_M\approx0, T, \delta T/M) = -\beta^{-1} M^{-1}\kappa_0 \delta
T = -(E_L/E_\mathrm{J}) \kappa_0 \delta T$ is independent of $M$, due to the
cancellation of the factors $M$ and $\beta$ occuring in $\kappa_{0M}$ and $\delta T_M$.  Moreover, as $E_L
\ll E_\mathrm{J}$, the heat current through the superinductance is negligible
compared to the one through the \textit{black sheep}, which is proportional to
$-\kappa \delta T$. We therefore completely neglect the heat current through the
superinductance in the following.

\subsection{Sensitivity of the heat current to the state of the fluxonium qubit}\label{SmallDeltaF}

\begin{figure}[tb]
\centering
\includegraphics[width=.7\linewidth]{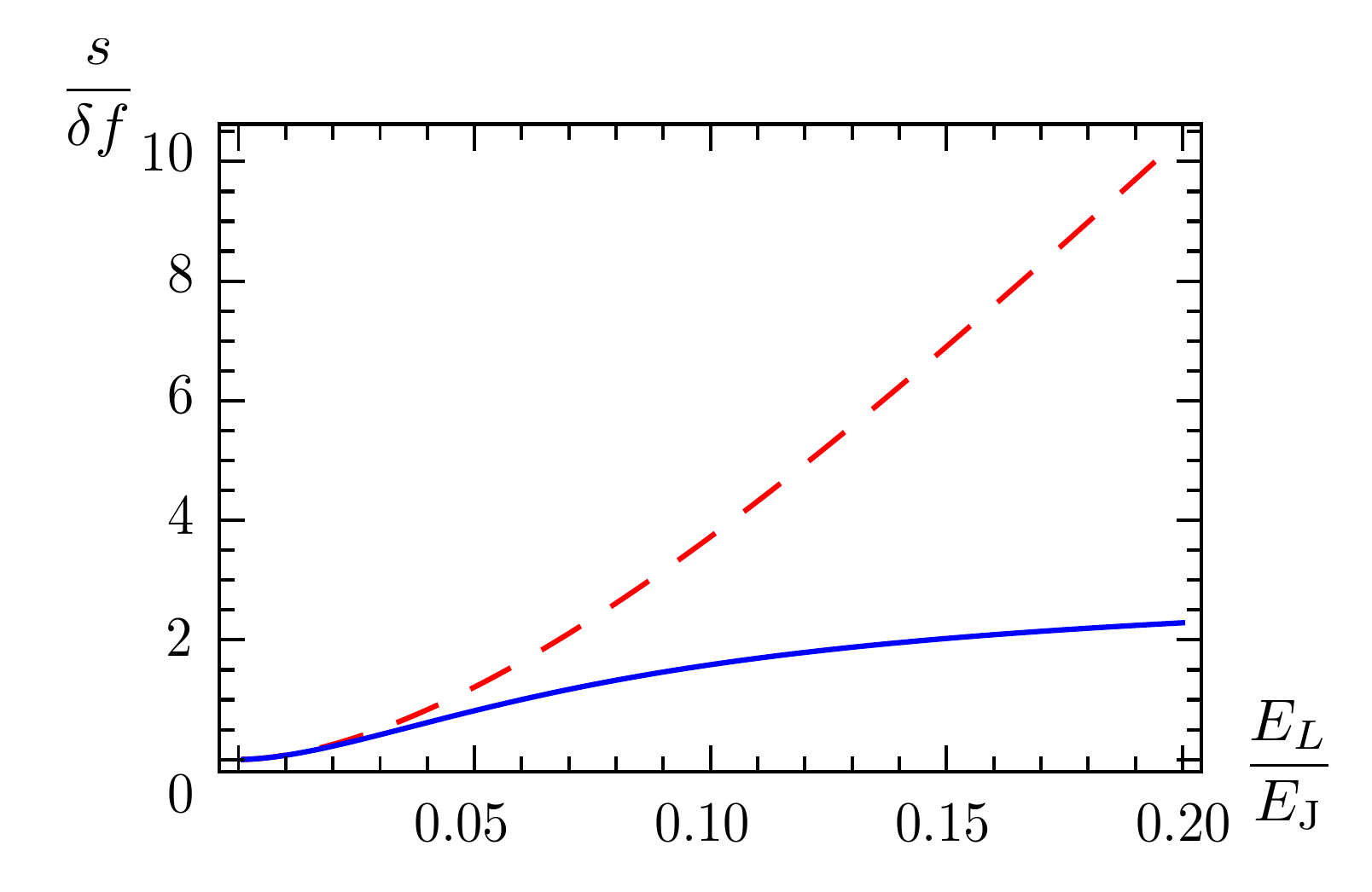}
\caption{Sensitivity as a function of $E_L/E_\mathrm{J}$ for $D=10^{-2}$ and
$k_\mathrm{B}T=0.1 \Delta$. The full result of equation~(\ref{s}) (full blue line) is compared to its
approximation for $\delta f,E_{L}/E_\mathrm{J}\rightarrow 0$ of
equation~(\ref{eqsensitivity3}) (red dashed line). }\label{SensitivityApprox}
\end{figure}

Having found that the sensitivity is independent of the specific realization of the superinductance, we here present an analytical expression for the sensitivity
$s$ in the regime $\delta f \ll 1$ and $E_L \ll E_\mathrm{J}$, which is relevant for the
fluxonium qubit. We have seen in section \ref{qubit_states} that $\varphi_R =
\varphi^* - 2\pi (1-\pi E_L/E_\mathrm{J}) \delta f$ and $\varphi_L = -\varphi^* - 2\pi
(1-\pi E_L/E_\mathrm{J}) \delta f$ with $\varphi^* = \pi (1- E_L/E_\mathrm{J})$. Evaluating $s$
of equation \ref{s} to first order in $\delta f$ and leading order in
$E_L/E_\mathrm{J}$ yields the final result
\begin{eqnarray}\label{eqsensitivity3}
  s\approx \frac{2\pi^2 \kappa_1}{\kappa_0}  \bigl|\ln (E_L/E_\mathrm{J}) \bigr|
  \frac{E_L^2}{E_\mathrm{J}^2} \delta f.
\end{eqnarray}
This expression shows that the sensitivity depends quadratically on the ratio
$E_L/E_\mathrm{J}$ with logarithmic corrections. Thus, the heat current in the fluxonium qubit is less sensitive to the qubit state than it is the case for the Delft qubit, due to the
smaller ratio $E_L/E_\mathrm{J}$. 
This comparison can be easily performed, since a description of both qubits is possible following equation (\ref{Hamiltonian}), with the effective qubit parameters given in table \ref{table}.
In what follows, we show that the fluxonium qubit therefore enjoys an increased protection with respect to
quasiparticle processes as compared to the Delft qubit.  In
figure~\ref{SensitivityApprox}, we have plotted the approximate result of the sensitivity 
 of equation \ref{eqsensitivity3} exhibiting a quadratic dependence for small $E_L/E_\mathrm{J}$; A comparison to the exact expression \ref{s}
indicates that the approximation is valid for $E_L/E_\mathrm{J} \lesssim 0.05$.

\section{Dephasing time}\label{sec_dephasing}

In a recent work \cite{Spilla14}, we have demonstrated that the sensitivity of
the heat current in a flux qubit (in the Delft qubit design) leads to a
dephasing of the qubit. The reason for this dephasing is the fact that for
non-zero sensitivities the tunnelling probabilities  of quasiparticles depends
on the state of the qubit; hence quasiparticles which tunnel through the
Josephson junction, can dephase the qubit. Importantly, the heat current incorporates both the phase-dependent quasiparticle transmission probabilities through the junction as well as the quasiparticle distribution functions. This forms the basis for our argument that the heat current captures the relevant properties leading to qubit dephasing due to quasiparticle tunneling. We therefore propose to investigate this  transport property, in order to access quasiparticle dephasing on a phenomenological level. We gain additional confidence in our results for the qubit dephasing obtained from this phenomenological approach by noticing that in the regime of large temperatures (where the cut-off due to the Andreev bound state becomes unimportant) our approach reproduces the perturbative results from the microscopic model \cite{Catelani12}.

Following reference~\cite{Spilla14}, see also \ref{app:der},  we can derive
the expression
\begin{eqnarray}\label{eq:deph}
 \tau^{-1}_\phi
 =\frac{2\Delta^4\pi^4}{e^2R } \left(\frac{E_L\delta f}{E_\mathrm{J}}\right)^2\!\!
\int_{\Delta}^{\infty}d\omega\,\frac{\omega^2-\Delta^2}{(\omega^2
  -\omega_\mathrm{c}^*{}^2)^2 (\omega^2 + \Delta^2)
  \cosh^{2}(\omega/ 2k_\mathrm{B} T)} 
\end{eqnarray}
for the inverse dephasing time, valid to lowest order $E_L/E_\mathrm{J}$ and $\delta f$; here,
$\omega_\mathrm{c}^*=\Delta (1-D \sin^2 \frac{\varphi^*}{2})^{1/2}$ is the
bound state energy at the phase difference $\varphi^*$ at $\delta f=0$. Note
that at the sweet spot the heat currents are equivalent for the two qubit
states which results in the vanishing of the dephasing rate, up to exponentially small corrections in $E_\mathrm{J}/E_\mathrm{C}$, which are not considered in this paper.

Owing to our phenomenological approach, the dephasing time can be directly
brought into contact with the sensitivity of the heat currents flowing in the
device to the qubit state. In order to estimate this link, we consider the low
temperature regime, $T,\delta T\ll \Delta/k_\mathrm{B}$, and write down the
product between dephasing time and the difference in heat currents in the two
qubit states, $\delta \dot{Q}$.  This function gives us an idea about the
energy which is transferred by the \textit{difference} of heat currents in the
two qubit states in the time, which the qubit needs to dephase. It turns out
that to lowest order in
$E_L/E_\mathrm{J}$ and $\delta f$, we obtain the simple result
\begin{equation}\label{3}
\tau_\phi \delta\dot Q \approx
\frac{ \Delta^2\delta T}{4k_\mathrm{B}T^2 \delta f}.
\end{equation}
The full expressions are presented in \ref{app:link}.  Equation \ref{3} shows
that the value of $\tau_\phi \delta\dot Q $ only depends on the detuning
$\delta f$ from the sweet spot and is otherwise independent of any of the
qubit parameters. The additional parameters $\Delta, T, \delta T$ entering the
expression describe the heat current due to the flow of the quasiparticles.

Starting from equation \ref{3}, we only need to find  $\delta \dot Q$ in order
to obtain the dephasing time $\tau_\phi$ that limits the performance of the
superconducting qubit.  In the limit $\delta f$ small, the two states $|\psi_\mathrm{L}\rangle$ and
$|\psi_\mathrm{R}\rangle$ have a similar heat current with $\dot Q_\mathrm{L} + \dot Q_\mathrm{R} =2 \dot
Q(\varphi=\varphi^*)$, such that we obtain the expression $\delta \dot Q= 2 s |\dot
Q(\varphi=\varphi^*)|$. Additionally, in the limit $E_L\ll E_\mathrm{J}$, $\varphi^*$ is
close to $\pi$ and thus $\dot Q(\varphi=\varphi^*)=-\kappa_0 \delta T$ which
leads to the final result
 \begin{equation}\label{5}
   \tau_\phi\approx \frac{\Delta^2}{8\kappa_0 k_\mathrm{B}T^2\delta f}
   \frac1s.
\end{equation}
We see that as expected the dephasing time is inversely proportional to the
sensitivity of the heat current that is flowing through the structure to the state of
the qubit. As we have seen before, $s$ is proportional to $(E_L/E_\mathrm{J})^2$ (up to
logarithmic corrections); consequently the fluxonium qubit has the benefit of a larger
dephasing time than the Delft qubit. Different from the dephasing caused by charge and flux noise, the
inverse dephasing time due to the transport of quasiparticles is not exponentially
suppressed in the fluxonium regime $E_L \ll E_\mathrm{J}$. However, due to the fact
that $\kappa_0 \propto \exp(-\Delta/k_\mathrm{B} T)$, an exponential improvement of the dephasing time limited by quasiparticle tunneling can be reached by lowering the \emph{temperature}. This is expected since it corresponds to decreasing the quasiparticle occupation.

\begin{figure}
\centering
\includegraphics[width=0.7\linewidth]{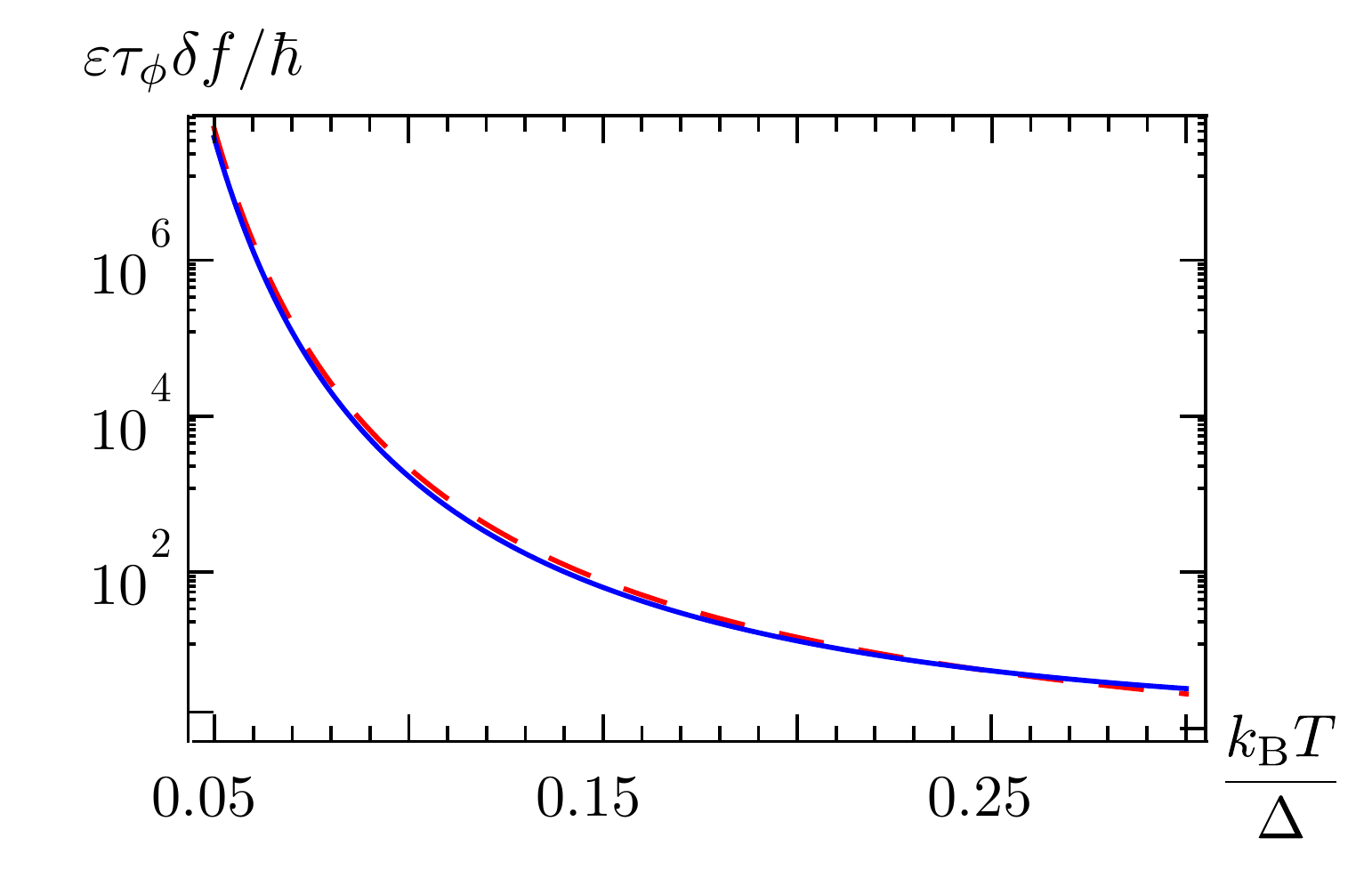}
\caption{Dephasing time, $\tau_\phi$, multiplied by the level splitting
of the fluxonium $\varepsilon$ and by $\delta f$  as a function of
temperature (full blue line). The transmission of the
junction is assumed to be $D=10^{-2}$, $E_\mathrm{L}/E_\mathrm{J}=6\times
10^{-2}$ and the number of channels contributing to transport $N\approx10^{3}$.  The red dashed line shows a fit $d\times\exp(\Delta/k_\mathrm{B}T)$
with $d=10^{-1}$.}
\label{fig_DephBohrDf}
\end{figure}

In order to make a quantitive statement relevant for applications, it is useful to introduce the
dimensionless number $\varepsilon \tau_\phi/\hbar$ with $ \varepsilon$ the
level splitting of the qubit, rather than looking at the dephasing time only. In particular, since the inverse level splitting, $\varepsilon^{-1}$, yields a measure of the time of a qubit operation, the parameter $\varepsilon \tau_\phi/\hbar$ indicates the
typical number of single qubit gates which can be performed before coherence
is lost. We numerically determine the  level splitting  from the difference of
the energies of the ground and excited state from the Hamiltonian given in
equation~\ref{Hamiltonian}.  In figure~\ref{fig_DephBohrDf}, we show a plot of
the function $\varepsilon\tau_\phi\,\delta f/\hbar$. Since with small changes in
$f$, the level splitting increases linearly with $\delta f$, while $\tau_\phi$
is proportional to $\delta f^{-2}$, we choose to multiply the parameter of
interest with $\delta f$ in order to get a function which is essentially
independent of $\delta f$.  Figure~\ref{fig_DephBohrDf} shows a
logarithmic plot of $\varepsilon\tau_\phi\,\delta f/\hbar$ for small values of the
temperature, $T\ll \Delta/k_\mathrm{B}$.  The ratio between the two
time-scales, $\tau_\phi$ and $\hbar/\varepsilon$, occurs to be approximately exponentially suppressed with
increasing temperature. This can be seen by using equations \ref{eqsensitivity3} and \ref{5}, where we see that $\tau_{\phi}\propto T^{-2}\kappa_1(T)^{-1}$. Moreover, from the definitions of $\kappa_1(T)$ given in equation \ref{ks}, for small temperatures we have $T^2\kappa_1(T)\propto\exp(-\Delta/k_\mathrm{B}T)$.

Finally, figure \ref{fig_DephBohrDf} shows that for
small values of the temperature it is possible to perform a great number of
operations on the qubit state before they become unreliable due to dephasing
of the two-level system. For example for $\delta f=10^{-2}$ and $ k_\mathrm{B}
T < 0.15 \Delta$, we have that $\varepsilon\tau_\phi/\hbar > 10^4$.

\section{Conclusion}

In this paper, we have studied the impact of the phase sensitivity of the
quasiparticle transport on the coherence properties of the fluxonium qubit.
Using a phenomenological approach, based on the study of heat currents carried by quasiparticles and the associated heat conductance, we have shown that the dephasing time is
inversely proportional to the sensitivity, a quantity describing to which extent possible
heat currents flowing in the device depend on the state of the qubit.  We have
shown that the sensitivity of the heat current to the qubit state depends
quadratically on the ratio $E_L/E_\mathrm{J}$ of the characteristic energies of the
fluxonium qubit but not on the number of junctions with which the
superinductance, a relevant ingredient of the fluxonium qubit, is realised.  The independence of the number of
array-junctions can qualitatively be traced back to the fact that, at small
temperature gradients, the heat current in the arms of the loop constituting
the fluxonium qubit is mainly given by the heat current in the so-called \textit{black
sheep} junction, which does not depend on $M$. The fact that the sensitivity can be reduced
by lowering the ratio $E_L/E_\mathrm{J}$ has the important result that the fluxonium qubit is
less affected by dephasing due to quasiparticle tunnelling through the
Josephson junction when compared to the Delft qubit design. We furthermore find that the dephasing mechanism is
exponentially suppressed with temperature due to its origin in quasiparticle tunneling. However, we have shown that at
moderately low temperatures the  resulting dephasing time is demonstrated to
be large enough to easily allow an excess of $10^4$ operations before the
qubit dephases. 

\ack We thank G.~Catelani and R.~Fazio  for fruitful discussions.  FH
acknowledges financial support from the Alexander von Humboldt foundation. JS acknowledges financial support from the Knut and Alice Wallenberg foundation through the Wallenberg Academy Fellows program and from the Swedish VR.

\appendix

\section{Evaluation of the linear-response coefficients for weak tunnel
coupling}\label{app:weak}

We aim to get an understanding of the phase dependence of the linear response coefficient $\kappa$ of the heat current in the weak tunnel coupling regime, $D\ll1$ as given in equation~\ref{thermal current},
\begin{eqnarray}\label{coeff_full}
\kappa=\frac{1}{2e^2R
  k_\mathrm{B}T^2}\int_{\Delta}^{\infty}d\omega\,\frac{\omega^2}{
    \cosh^2{(\omega/2k_\mathrm{B}T)}}\frac{\omega^2-\Delta^2}{(\omega^2-\omega_\mathrm{c}^2)^2}(\omega^2-\Delta^2\cos{\phi}) ,
\end{eqnarray}
by rewriting it in terms of the expression given in equation~\ref{ks}. In this
appendix, we outline the procedure, which we apply to obtain the coefficients given in equation~\ref{ks}. 

As a starting point, it is useful to evaluate the logarithmic divergence occuring in the linear response coefficient shown in equation~\ref{coeff_full}. Since the divergence stems from values of $\omega$ in the vicinity of the superconducting gap $\Delta$, this logarithmic divergence can conveniently be extracted by setting  $\omega\approx\Delta$ in all contributions of equation~\ref{coeff_full} with a smooth dependence on $\omega$ in the vicinity of $\Delta$. This consideration leads us to determine the integral
\begin{eqnarray}
\kappa_\mathrm{div} & \equiv & \frac{\Delta^3\sin^2\frac\phi2}{e^2R
  k_\mathrm{B}T^2\cosh^2\left(\Delta/2k_\mathrm{B}T\right)}\int_{\Delta}^{\sqrt{3}\Delta}d\omega\,\frac{\omega}{\omega^2-\Delta^2\left(1-D\sin^2\frac{\phi}{2}\right)}\\
  & = & -\frac{\Delta^3\sin^2\frac\phi2}{2e^2R
  k_\mathrm{B}T^2\cosh^2\left(\Delta/2k_\mathrm{B}T\right)}\ln\left(D\sin^2\frac\phi2\right)\nonumber.
\end{eqnarray}
The latter is the only divergent contribution in equation~\ref{coeff_full}. With $R^{-1}\propto D$, we find that the leading contribution to this term is of the order $D\ln D$. The remaining part of the integral can safely be expanded for small $D\sin^2\frac\phi2$. We find the remaining part by simply substracting the logarithmic divergence from the full coefficient $\kappa_\mathrm{rem}\equiv\kappa-\kappa_\mathrm{div}$, leading to 
\begin{eqnarray}\label{remaining}
 \kappa_\mathrm{rem} & = &\frac{1}{2e^2R
  k_\mathrm{B}T^2}\int_{\Delta}^{\infty}d\omega\,\frac{\omega^2}{
    \cosh^2{(\omega/2k_\mathrm{B}T)}}\frac{\omega^2-\Delta^2}{(\omega^2-\omega_\mathrm{c}^2)^2}(\omega^2-\Delta^2\cos{\phi})\\
  &&  -\frac{\Delta^3\sin^2\frac\phi2}{e^2R
  k_\mathrm{B}T^2\cosh^2\left(\Delta/2k_\mathrm{B}T\right)}\int_{\Delta}^{\sqrt{3}\Delta}d\omega\,\frac{\omega}{\omega^2-\Delta^2\left(1-D\sin^2\frac{\phi}{2}\right)}
    \nonumber
  \end{eqnarray}
 Indeed, the expansion of equation~\ref{remaining} to linear order in $D$ shows no divergent behavior anymore. We find
 \begin{eqnarray}
\fl \kappa_\mathrm{rem}  = & \frac1{2e^2Rk_\mathrm{B}T}\int_\Delta^\infty d\omega\frac{\omega^2}{\cosh^2\left(\omega/2k_\mathrm{B}T\right)}\nonumber\\
\fl &+\frac{\Delta^2}{2e^2Rk_\mathrm{B}T}\int_{\sqrt{3}\Delta}^\infty d\omega\frac{\omega^2}{\cosh^2\left(\omega/2k_\mathrm{B}T\right)}\frac{\sin^2\frac\phi2}{\omega^2-\Delta^2}\nonumber\\
\fl & +\frac{\Delta^2}{2e^2Rk_\mathrm{B}T}\int_\Delta^{\sqrt{3}\Delta}d\omega\frac{\omega}{\omega^2-\Delta^2}\left[\frac{\omega}{\cosh^2\left(\omega/2k_\mathrm{B}T\right)}
 - \frac{\Delta}{\cosh^2\left(\Delta/2k_\mathrm{B}T\right)}
 \right]\sin^2\frac\phi2 \nonumber
 \end{eqnarray} 
This expression, together with the contribution from the logarithmic divergence $\kappa_\mathrm{div}$, yields the results presented in equations ~\ref{zhao} and \ref{ks} in the main text.

\section{Derivation of the dephasing time}\label{app:der}

In order to investigate the impact of possible temperature gradients on the dephasing of the fluxonium qubit, we follow the lines of reference~\cite{Spilla14}. We therefore use a model Hamiltonian, $H_\mathrm{mod}=H_0+H_\mathrm{I}$, where $H_0$ describes the qubit as a two-level system with states $|\psi_\mathrm{L}\rangle$ and $|\psi_\mathrm{R}\rangle$ and normal quasi-particle reservoirs at different temperatures. The coupling between them is given by $H_\mathrm{I}$. More specifically, we have
\begin{eqnarray}\label{eq:toy}
\fl H_\mathrm{mod} & = & -\frac{\varepsilon}{2}\tau^3+\sum_{l=1,2}\sum_{k,\sigma}(\varepsilon_{l,k} -\mu_l) c_{l,k\sigma}^{\dagger}c_{l,k\sigma}
^{\vphantom\dag} + \sum_{k,q,\sigma} \left[
 ( V_0\tau^0
 + V_3 \tau^3
 )  c_{1,k \sigma}^\dagger c_{2,q\sigma}^{\vphantom\dag} 
 + \mathrm{H.c.} \right]
\end{eqnarray}
The matrices $\tau^j, j=0,3$ are Pauli matrices in the qubit space. The level  splitting between the qubit states is given by $\varepsilon$; coupling between them is supposed to be weak and is neglected here. In the reservoirs, $l$ $(l=1,2)$, the creation (annihilation) operators of particles with momentum $k$ and spin $\sigma$ are given by $c^{\dagger}_{l,k\sigma}(c_{l,k\sigma})$. It is the state-dependent coupling between qubit and reservoirs occurring in the interaction part of the Hamiltonian, $V_{0/3}  = (V_\mathrm{R} \pm V_\mathrm{L})/2$, together with the density of states of the reservoirs, which takes account for the phase dependence (and hence for the dependence on the fluxonium states) of the heat current due to the superconductors.

We recover the state-dependent qubit-reservoir coupling and the density of states by comparing the heat current obtained from the model Hamiltonian in the linear response regime
\begin{equation}\label{eq:thermal}
\dot Q^\mathrm{mod}_\mathrm{R/L} 
= - \frac{\pi \delta T}{ 4 \hbar k_B T^2} 
\int_{\Delta}^{\infty} d\omega\, \frac{\omega^2}{\cosh^2(\omega/2 k_\mathrm{B} T)} 
V^2_\mathrm{R/L}
   n^\mathrm{R/L}_{1} n^\mathrm{R/L}_{2}, 
\end{equation}
to the one calculated starting from equations~(\ref{eq_heatgeneral}) and (\ref{eq_totalheat}) for the \textit{black sheep}. Note that we here neglect the effect of the heat current through the array junctions, since no relevant dependence on the qubit state occurs there. 
In order to extract the local density of states of the continuum states above the gap, we use the relation~\cite{Zhao04}
\begin{eqnarray}\label{eq:v_and_rho2}
   n^\mathrm{R/L}_{l}(\omega)
   &= n_l^0 \langle \varphi_\mathrm{R/L}|
   \frac{|\omega|(\omega^2-\Delta^2)^{1/2}}{\omega^2- \omega_\mathrm{c}^2}
  |\varphi_\mathrm{R/L}\rangle\nonumber\\
  &= n_l^0  \frac{|\omega|(\omega^2-\Delta^2)^{1/2}}{\omega^2- \Delta^2(1- D
    \sin^2 \frac{\varphi_\mathrm{R/L}}2)}
.\nonumber
\end{eqnarray}
Here,
$n^0_l$ is the normal conducting density of states including spin.
Using this form for the density of states we have the following expression for the tunnelling matrix elements  in the tunnelling regime, $D\ll1$,
\begin{eqnarray}\label{eq:v_and_rho}
V^2_\mathrm{R/L}(\omega)=V^2_{12}\left(1-\frac{\Delta^2}{\omega^2} \cos{\varphi_\mathrm{R/L}}\right).
\end{eqnarray}
Here, $V_{12}$ is the tunnelling amplitude of the junction figuring as the
\textit{black sheep} in the fluxonium qubit. It is linked to the normal state
resistance by $R= \hbar/(\pi e^2 n_1^0 n_2^0 V_{12}^2)$. 

With the help of this simplified model, we proceed to study the dynamics of the qubit state. Starting from the density matrix of the full system consisting of the qubit coupled to reservoirs, we trace out the reservoir degrees of freedom and write down a master equation for the reduced density matrix of the qubit, $\rho(t)$.
It is helpful to rewrite the interaction part of the Hamiltonian as $H_\mathrm{I}=P_\mathrm{R} B_\mathrm{R}+P_\mathrm{L} B_\mathrm{L}$ with the projectors on the qubit states, $\alpha=\mathrm{R,L}$,
\begin{eqnarray}
P_\alpha=|\psi_\alpha\rangle\langle \psi_\alpha|,
\quad \quad B_\alpha=V_\alpha\sum_{k,q,\sigma} c_{1,k \sigma}^\dagger
c_{2,q\sigma} + \mathrm{H.c.} 
\end{eqnarray}
Following a standard procedure, see for example Ref.~\cite{Petruccione}, the master equation for the density matrix of the qubit then takes the form
\begin{eqnarray}\label{MasterEq}
\dot{\rho}(t)=-\frac{i}{\hbar}[H_\mathrm{S},\rho(t)]
+\sum_{\alpha,\beta=\mathrm{R,L}}\gamma_{\alpha\beta}\Bigl(P_\beta\rho(t)P_\alpha-\frac{1}{2}\{P_\alpha P_\beta,\rho(t)\}\Bigr)\nonumber
\end{eqnarray}
with the transition rates between qubit states $\alpha,\beta=\mathrm{R,L}$,
\begin{equation}
\gamma_{\alpha\beta}=\frac{1}{2}\int\langle\{B_\alpha(t),B_\beta(0)\}\rangle dt.
\end{equation}
The relaxation behavior of the qubit becomes particularly clear when rewriting the master equation in terms of a Pauli rate equation for the pseudo-spin states of the qubit, $\boldsymbol{S}(t) =\mathrm{Tr}[ \rho(t)
\boldsymbol{\tau} ]=[\rho_\mathrm{LR}(t)+\rho_\mathrm{RL}(t),
i\rho_\mathrm{LR}(t)-i\rho_\mathrm{RL}(t),
\rho_\mathrm{LL}(t)-\rho_\mathrm{RR}(t)]^\mathrm{T}$. We obtain from equation (\ref{MasterEq}),
\begin{equation}\label{eq_Pauli}
\dot{\boldsymbol{S}}(t)=
\boldsymbol{S}(t) \times \boldsymbol{h}-
\tau_\phi^{-1}(S_1(t),S_2(t),0)^\mathrm{T}
\end{equation}
The Pauli rate equation~(\ref{eq_Pauli}) contains a precession of the pseudospin around a pseudo-magnetic field, $\boldsymbol{h}= (0,0,\varepsilon/\hbar)^\mathrm{T}$, determined by the level splitting between qubit states. 
Most importantly, there is also a relaxation of the coherences of the reduced
density matrix with the dephasing rate $\tau_\phi^{-1}$, given by
\begin{equation}
 \tau_\phi^{-1}=\frac{1}{2\hbar}(\gamma_\mathrm{RR}-2\gamma_\mathrm{RL}+\gamma_\mathrm{LL}).
\end{equation}
It is found to have the explicit form,
\begin{eqnarray}
  \fl \tau_\phi^{-1}&=\frac{\pi}{2\hbar}
  \int_{\Delta}^\infty
  d\omega \Bigr[V_\mathrm{R}(n^\mathrm{R}_{1}n^\mathrm{R}_{2})^{1/2}
  -V_\mathrm{L} (n^\mathrm{L}_{1}n^\mathrm{L}_{2})^{1/2}\Bigr]^2
\cosh^{-2}(\omega/ 2k_\mathrm{B} T)\\
\fl &=\frac{1}{2e^2R}
\int_{\Delta}^{\infty}d\omega\,\frac{\omega^2-\Delta^2}{(\omega^2
  -\omega_\mathrm{c}^*{}^2)^2
  \cosh^{2}(\omega/ 2k_\mathrm{B} T)}
 \Bigl( \sqrt{\omega^2-\Delta^2\cos{\varphi_\mathrm{L} }} -
\sqrt{\omega^2-\Delta^2\cos{\varphi_\mathrm{R} }}
 \Bigr)^2 \nonumber
 \end{eqnarray}
 where we introduced $\omega_\mathrm{c}^*=\Delta (1-D \sin^2 \frac{\varphi^*}{2})^{1/2}$.

 \section{Link between heat currents and dephasing time}\label{app:link}

As discussed in the main text, the dephasing time can be directly brought into
connection with the heat current flowing through the qubit due to a finite
temperature gradient and its sensitivity to the qubit state. In order to
estimate this link, we consider the linear response regime $\delta T \ll T$  and write down the product between dephasing time and the difference in heat currents in the two qubit states, $\delta \dot{Q}$,
\begin{eqnarray}\label{ratio}
  \fl
\tau_\phi \delta\dot Q =\frac{\Delta^2 \delta T\int_{\Delta}^{\infty} d\omega\,\omega^2
\frac{\omega^2-\Delta^2}{(\omega^2-\omega_\mathrm{c}^*{}^2)^2}
  \bigl|\cos{\varphi_\mathrm{L} } -\cos{\varphi_\mathrm{R} }\bigr|
\cosh^{-2}\frac{\omega}{2k_\mathrm{B}T}}{2k_\mathrm{B}T^2\!\!\int_{\Delta}^\infty
d\omega \frac{\omega^2-\Delta^2}{(\omega^2-\omega_\mathrm{c}^*{}^2)^2}
   \Bigl( \sqrt{\omega^2-\Delta^2\cos{\varphi_\mathrm{L} }}
   -\sqrt{\omega^2-\Delta^2\cos{\varphi_\mathrm{R} }}\Bigr)^2 \!\!\cosh^{-2}\frac{\omega}{2k_\mathrm{B}T}
}.
\end{eqnarray}
As we noticed before, the main contribution to these integrals stems from contributions of $\omega$ close to the superconducting gap $\Delta$. We hence introduce $\omega\approx\Delta$ in those factors which are smooth functions of $\omega$ in the vicinity of $\Delta$ 
\begin{eqnarray}\label{ratio1}
  \fl
\tau_\phi \delta\dot Q &=\frac{\Delta^2 \delta T\int_{\Delta}^{\infty}
d\omega\,\Delta^2
\frac{\omega^2-\Delta^2}{(\omega^2-\omega_\mathrm{c}^*{}^2)^2}
  \bigl|\cos{\varphi_\mathrm{L} } -\cos{\varphi_\mathrm{R} }\bigr|
\cosh^{-2}\frac{\omega}{2k_\mathrm{B}T}}{2k_\mathrm{B}T^2\!\!\int_{\Delta}^\infty
d\omega \frac{\omega^2-\Delta^2}{(\omega^2-\omega_\mathrm{c}^*{}^2)^2}
   \Bigl( \sqrt{\Delta^2-\Delta^2\cos{\varphi_\mathrm{L} }}
   -\sqrt{\Delta^2-\Delta^2\cos{\varphi_\mathrm{R} }}\Bigr)^2 \!\!\cosh^{-2}\frac{\omega}{2k_\mathrm{B}T}
} 
\end{eqnarray}
and observe that this leads to a cancellation of the integral terms in
numerator and denominator of equation~\ref{ratio1}. This underlines the close connection between the heat current sensitivity to the qubit state and the occurring dephasing mechanism. Further simplifying we find
\begin{eqnarray}
\tau_\phi \delta\dot Q &= \frac{\Delta^2 \delta T}{4k_\mathrm{B}T^2}\frac{  \bigl|\cos{\varphi_\mathrm{L} } -\cos{\varphi_\mathrm{R} }\bigr|
}{   ( |\sin \frac{\varphi_\mathrm{L}}2|    -
| \sin \frac{\varphi_\mathrm{R}}2  | )^2 }  .
\end{eqnarray}
In lowest order in $E_L/E_\mathrm{J}$ and $\delta f$, this leads to the result presented in equation~\ref{3} in the main text.
%
\section*{Bibliography}

\end{document}